\documentclass[12pt]{article}
\usepackage{graphicx}
\usepackage[dvips]{color}
\definecolor{violet}{rgb}{0.6,0.5,1.0}
\definecolor{dgreen}{rgb}{0.0,0.7,0.0}
\definecolor{orange}{rgb}{1.0,0.45,0.4}
\begin{document}

\centerline{\large \bf 
New determinations of gamma-ray line intensities of the E$_p$ = 550 keV}
\centerline
{\bf and E$_p$ = 1747 keV resonances of the $^{13}$C(p,$\gamma$)$^{14}$N reaction}
\vspace{1 cm}

\noindent
J. Kiener$^{1,a}$, M. Gros$^2$, V. Tatischeff$^1$, D. Atti\'e$^2$, I.
Bailly$^3$, A. Bauchet$^1$, C. Chapuis$^2$, B. Cordier$^2$, I.
Deloncle$^1$, M.G. Porquet$^1$, S. Schanne$^2$, N. de S\'er\'eville$^1$, 
G. Tauzin$^2$
\vspace{0.5 cm}

\noindent
\scriptsize
\centerline {(1) CSNSM, CNRS-IN2P3 et Universit\'e Paris-Sud,
F-91405 Campus Orsay, France}
\centerline{(2) DSM/DAPNIA/Service d'Astrophysique, CEA Saclay, 
F-91191 Gif-sur-Yvette, France}
\centerline{(3) CEA/DIF/DPTA/SP2A/Laboratoire des acc\'el\'erateurs
\'electrostatiques, BP 12, F-91680 Bruy\`eres-le-Ch\^{a}tel, France}
\centerline{(a) Corresponding author, e-mail: kiener@csnsm.in2p3.fr}
\vspace{0.5 cm}

Gamma-ray angular distributions for the
resonances at E$_p$ = 550 keV and 1747
keV of the radiative capture reaction $^{13}$C(p,$\gamma$)$^{14}$N have been
measured, using intense proton beams on isotopically pure $^{13}$C targets.
Experimental gamma-ray spectra were obtained with three
HP-Germanium detectors at four angles for E$_p$ = 550 keV and six
angles for E$_p$ = 1747 keV in the range of
0$^{\circ}$ to  90$^{\circ}$ with
respect to the proton beam. From the data, relative intensities for the
strongest transitions were
extracted with an accuracy of typically five percent, making these resonances
new useful gamma-ray standards for efficiency calibration in the energy
range from E$_{\gamma}$ = 1.6 to 9 MeV. Gamma-ray branching ratios were 
obtained 
for several levels of $^{14}$N and are compared with literature values.
\vspace{1.2 cm}

\normalsize
\section{Introduction}
Detection systems for nuclear gamma-ray lines often need to cover an
important part of the whole possible energy range of nuclear
transitions from some 10 keV to about 15 MeV. This is true
in nuclear spectroscopy, nuclear reaction analysis and gamma-ray
astronomy. Energy and efficiency calibration can generally be assured by
standard radioactive sources up to 3.5 MeV. Above that
energy, one has to use other sources like gamma rays
from nuclear reactions. There, the most used reaction is probably
the well-studied E$_p$ = 992 keV resonance of the radiative capture reaction
$^{27}$Al(p,$\gamma$)$^{28}$Si. It offers a great variety of gamma-ray
lines from 1.5 MeV to 10.7 MeV with known relative intensities 
\cite{Anttila}. 

This resonance has, however, no strong lines in
some relatively large portions of the above cited energy range. Another
slight drawback may be the moderate yield of about 10$^{-9}$
gamma rays per proton. The two resonances at E$_p$ = 550 keV and 1747 keV
of the radiative capture reaction  $^{13}$C(p,$\gamma$)$^{14}$N offer a
similar coverage of the energy range and a factor of about five higher
gamma-ray yields. This makes them interesting
complementary or alternative gamma-ray sources for energy and 
efficiency calibrations.   

These two $^{13}$C(p,$\gamma$) resonances have been used for the ground
calibration of the gamma-ray spectrometer SPI  of the INTEGRAL satellite
in the energy range E$_{\gamma}$ = 1.6 to 9 MeV \cite{Schanne}.
In the analysis of these calibration runs, it became clear that the
relative line intensities calculated from the published branching
ratios of the involved $^{14}$N levels needed reinvestigation \cite{GKT}.
We therefore decided to determine experimentally the relative intensities
for the strongest lines of both resonances aiming at accuracies of better
than ten per cent. This was achieved by the measurement of high-statistics 
spectra with HP-Ge detectors, for both resonances 
in the angular range of 0$^{\circ}$ to 90$^{\circ}$.

A particular effort was made at  45$^{\circ}$ with respect to the proton 
beam,  which was the angle of SPI 
during the ground calibration runs, where we obtained relative intensities
of the strongest lines with an accuracy of three to five 
per cent. In order to allow detector efficiency calibrations 
at any angular position, gamma-ray angular distributions for both resonances
were determined. For the 550 keV resonance, which is dominated by s-wave
capture with a small probable contribution of d-wave capture, gamma-ray line
intensities were measured at four angles. The 1747 keV resonance has p-wave
and f-wave capture contributions and data were taken at six angular positions 
to determine the gamma-ray angular distributions. 

Additionally, angle-integrated line intensites from Legendre
polynomial fits to the angular distribution data permitted
extraction of gamma-ray branching ratios for several levels of $^{14}$N.
Angle-integrated line intensities and branching ratios are then compared 
to recent literature values and discussed.
Finally, we propose new values for the gamma-ray branchings of the 
6.446, 8.062 and 9.172 MeV levels of $^{14}$N.  

\section{Experiment}

The experiment was done at the 3 MV Van-de-Graaff accelerator of the CEA at
Bruy\`eres-le-Ch\^{a}tel.  Proton beams of 100 to 250 $\mu$A intensity
were directed onto an isotopically pure $^{13}$C target of 129
$\mu$g/cm$^2$, evaporated on a water-cooled tantalum disk. 
The target chamber was isolated from the rest of the beam line and 
served as a Faraday cup
for the beam current integration. During the irradiation runs, the water
temperature was monitored to detect a possible degradation of the target or 
deviations of the proton beam.

Three coaxial HP-Ge detectors of 100-250 cm$^3$
were used for the detection of the gamma rays. Two of the
detectors were used for monitoring purposes, sitting at fixed positions 
at 45$^{\circ}$ to the left  (detector named ``D03'') and 
to the right (detector named ``P2F'') with respect to the proton beam.
D03 and P2F were placed at 1 m and 1.675 m distance from the target
position, respectively.
The third detector (named ``D02'') was used for the angular distribution
measurements and
took spectra at 0$^{\circ}$, 30$^{\circ}$, 60$^{\circ}$ and
90$^{\circ}$ for both resonances and additionally at 15$^{\circ}$ and
45$^{\circ}$ for the 1747 keV resonance, at a distance of 1 m from the target.
Positioning accuracy was estimated to be better than 5 mm for D02 and P2F,
and better than 1 cm for D03. Energy and efficiency calibration of the
detectors were done with natural background lines, an $^{88}$Y source and the
E$_p$ = 992 keV $^{27}$Al(p,$\gamma$) resonance. The energy resolutions (FWHM) 
of the detectors at 1836/9200 keV were 2.6/6.9 keV (D03), 
3.0/9.2 keV (P2F) and 2.4/7.2 keV (D02).

Before starting the long irradiation runs, both resonances were scanned in
steps of 5 to 10 keV to find the maximal gamma-ray yield. This is especially
important for the E$_p$ = 550 keV resonance, whose resonance energy and
width ($\Gamma_t$ $\approx$ 30 keV; this corresponds approximately to the
proton energy loss in the target) is relatively uncertain
(see e.g. \cite{Gal} and references therein).
The E$_p$ = 1747 keV resonance is a narrow resonance ($\Gamma_t$ = 122 eV)
and the gamma-ray yield reaches a plateau for E$_p$ above 1748 keV for the
target used in this experiment. The energy chosen was approximately in the
middle of the yield plateau.  The same was done for the
narrow E$_p$ = 992 keV $^{27}$Al(p,$\gamma$) resonance, which was used for
efficiency calibration of the detectors.
It was also checked, that the yield was independent of the beam current in 
the range 20 to 200 $\mu$A.

The long irradiation runs were done with beam currents of typically
250 $\mu$A for the 550 keV resonance,  200 $\mu$A for the 992 keV resonance,
and 100 $\mu$A for the 1747 keV resonance and lasted for at least three hours. 
For the strongest gamma-ray lines of each resonance in these runs, a minimum
of 25000, 60000 and 25000 events for the 9169 keV, 1779 keV and 8160 keV
lines, respectively, have been collected in the full-energy peak for
detector P2F. Detectors D03 and D02 were closer to the target and registered
on average about a factor of two more events. No target degradation was
observed during the experiment. 

Efficiency calibration of the three HP-Ge detectors was done with the
$^{27}$Al(p,$\gamma$) resonance and an intense $^{88}$Y source.
High-statistics spectra of the resonance were taken with the monitor
detectors at 45$^{\circ}$ and with D02 at
0$^{\circ}$, 60$^{\circ}$ and 90$^{\circ}$, to obtain effective efficiency
curves at these angles which include implicitely the energy- and 
angle-dependent effect of absorption in the target backing and the
support structure. Special attention was paid to
obtain sufficient statistics for all gamma rays of the
reaction, whose angular distribution were determined by Anttila et al. 
\cite{Anttila} and to have the same conditions in the irradiation runs
of this $^{27}$Al(p,$\gamma$)
resonance and the two $^{13}$C(p,$\gamma$) resonances.
The $^{88}$Y source has also been used to
determine the gamma-ray absorption in the tantalum backing, and the
support structure of the target, by taking
spectra with D02 at eight different angles in the range 0$^{\circ}$ to
110$^{\circ}$. 

\section{Data analysis}

Cumulative spectra of the different runs for the three detectors were
acquired with two independent commercially available 
acquisition systems with
automatic deadtime correction capabilities. Energy calibration was done 
with the natural background lines of
$^{40}$K and the decay chains of $^{226}$Ra and  $^{228}$Th
and some selected  $^{27}$Al(p,$\gamma$) lines. Extraction of 
the gamma-ray line intensities were entirely based on the 
full-energy peaks with
the exception of some background lines, as explained below.

Peak integration including background subtraction was done with the GF3 
code
from the RadWare software package \cite{Radf}. Several gamma-ray
lines of the $^{13}$C(p,$\gamma$) resonances are broad and some of them
have complex structures due to gamma-ray emission during the slowing down
process of the $^{14}$N in the target. In these cases, peak  
integration and background subtraction were guided by simulations of the
line profiles. Details about this procedure can be found in \cite{GKT}. 
An additional estimated uncertainty 
was linearly added to the statistical uncertainty in cases where the background
level was judged uncertain.

\subsection{Detector efficiencies}

We aimed at the extraction of relative line intensities for all
transitions with intensities above $\approx$ 1\% with respect to one
capture reaction for both $^{13}$C(p,$\gamma$) resonances. That comprises
gamma-ray lines with energies ranging from 1635 keV to 9200 keV,
meaning that the detector efficiency curves must be determined precisely
over the whole range. 
The $^{27}$Al(p,$\gamma$) resonance offers
gamma-ray lines between 1522 keV and 10763 keV whose intensities are well
studied. We used all gamma rays whose angular distribution have been
determined by Antilla et al. \cite{Anttila} plus the four gamma rays at 
1522, 1658, 7924 and 7933 keV to improve the energy coverage. 
For these 
last four gamma rays, we added a systematic uncertainty of twenty 
per cent to the intensity for possible anisotropies of the angular
distribution. 

The resonance lines completed with the
two lines of the $^{88}$Y source at 898 keV and 1836 keV, cover 
the desired energy range and allow also absolute efficiency calibrations. 
It permitted furthermore the study of two lines at 1340 keV 
and 728 keV, which have intensities $<$ 1\%. 
A simple linear interpolation of detector efficiency between
the data points however, would only result in a moderately good definition
of the efficiency curve in some energy domains. This holds notably for the
region between 6.5 MeV and 10.5 MeV where only two close gamma-ray lines
at 7924 and 7933 keV are available \cite{Anttila}.

The choice of the function for analytical fits to the data suffered equally
from the scarcety of $^{27}$Al(p,$\gamma$) gamma-ray lines
and the uncertainties of their intensities in some energy domains.
We performed therefore Monte-Carlo simulations of the gamma-ray response of
the three Ge detectors to find the functions which fit best their 
full-energy peak efficiency curves. The code GEANT was used for this purpose
\cite{GEANT}. As only the full-energy peak efficiency was looked for, we 
restricted the simulated detector geometry to
the Ge crystal and its housing. 
The simulations were done using the experimental
detector-source distances and crystal geometries, 
including the dead layer
due to the n$^+$ contact of the coaxial n-type HP-Ge, which was
very important for the high-energy response of the detectors. 
The detector crystal response was then simulated for monoenergetic
gamma rays between 728 keV and 11 MeV, whereby for each energy 2000
(or more)
full-energy peak events were used. 

The simulated relative full-energy peak
efficiencies were then multiplied with a transmission factor, which takes
into  account the absorption of the gamma rays in the target-support
structure and the air between the target and the detectors. The absorption 
in the target support was measured with the $^{88}$Y source and was found 
for emission angles between 0$^{\circ}$ and 45$^{\circ}$ to be in
agreement with calculations using the photon cross section database of
NIST \cite{NIST} and the properties of the support structure (i.e.
1 mm Ta, 9.5 mm water and 1 mm stainless steel).  
Finally, a power-law function of linear varying index with energy 
proved to give an excellent fit to both the experimental and the 
simulated relative efficiency data:

\begin{equation}
\epsilon(E_{\gamma}) ~ = ~ a \cdot E_{\gamma}^{~(b ~ + ~ c \cdot E_{\gamma})} 
\label{eq1}
\end{equation} 

, where E$_{\gamma}$ is the gamma-ray energy and 
a, b and c are constants which depend on the detector and its angular position.
The reduced $\chi^2$ for the fits of the experimental data ranged from 0.76 to
0.88 for detectors D03 and P2F at 45$^{\circ}$ and for D02 at 0$^{\circ}$,
while the deviation of the simulated data from the fits stayed well below
five per cent in the whole energy region.
For larger emission angles, besides the target-support structure there was
additional absorption  in the target chamber material and
absorption calculations became very complex. Comparison with simulated
efficiencies were therefore not done for these data. 
However, fits with the function
of eq.~\ref{eq1} still gave a very good description of the $^{27}$Al(p,$\gamma$)
data for detector D02 at
60$^{\circ}$ and 90$^{\circ}$ (reduced $\chi^2$ 0.81 and 1.16, respectively)
and were therefore also used for these efficiency curves. For D02 at
15$^{\circ}$, 30$^{\circ}$ and 45$^{\circ}$, the efficiency curve obtained
at 0$^{\circ}$ has been used, corrected for the additional absorption 
due to the non-normal incidence of the gamma rays in the target-support
structure.
Examples of efficiency data together with the power-law fits are shown in
fig. 1. The absolute normalization of the efficiency curves will be 
treated in the section on gamma-ray angular distributions.

\subsection{Relative line intensities}

The relative gamma-ray line intensities have been extracted at all measured 
angles
for both resonances. A special effort was made at 45$^{\circ}$, which was
the angle of the INTEGRAL spectrometer SPI during its calibration campaign
\cite{Schanne} \cite{GKT}. 
With both monitor detectors D03 and P2F, high-statistics spectra could
be obtained at 45$^{\circ}$ by
accumulation of several 3-hour runs. Additionally, one 3-hour
run for the 1747 keV resonance was taken with D02 at this angle. Relative line
intensities at the other angles have only been measured with detector 
D02, representing one 3-hour run at each angle 
with the conditions described in the section II.   

Relative intensities of eight lines were extracted for the 550 keV resonance.
The relatively strong lines from the 8062 $\rightarrow$ 4915 keV and
8062 $\rightarrow$ 5691 keV transitions were omitted, because they partially
overlap with escape peaks from other lines.
For the 1747 keV resonance, relative intensities for
a total of fourteen lines could be obtained. Partial level schemes of
$^{14}$N with the important gamma-ray transitions for both resonances are
displayed in fig. 2. 

Special attention had to be paid to several lines of the 1747 keV
resonance whose full-energy peaks were merged with background lines.
The 5105 keV line  is blended with the
second escape peak of the 6129 keV line from the
$^{19}$F(p,$\alpha$$\gamma$)$^{16}$O reaction. 
A similar problem affects the 1635 keV line whose full-energy peak
overlaps with the first escape peak of the 2143 keV line. 
The contribution of these components
were subtracted from the lines after determining the escape-peak to
full-energy-peak ratios for the three different detectors in the 
respective energy domains.  
A  background line of the
$^{228}$Th decay chain was blended with the 728 keV line. It has been 
subtracted by making use of the 2614 keV line intensity during the
irradiation runs and the observed 728 keV to 2614 keV line ratios in 
several radioactive background measurements performed during the experiment.

The uncertainty on the relative intensities comprises the statistical 
one with an eventual additional uncertainty due to uncertain background 
as explained
above, and an estimated uncertainty of five per cent for the efficiency
calibration, with the exception for the 728 keV line, where ten per cent
were taken. For the 5106, 1635 and 728 keV lines, uncertainties
due to the subtraction of the background lines were added.
All uncertainties were added quadratically. The 45$^{\circ}$ data are weighted
averages of the results of the three detectors. 
The relative 
intensities are presented in table~1 for the 550 keV resonance
and in table~2 for the 1747 keV resonance.

\subsection{Angular distributions}

Angular distributions were determined for both resonances. The 550 keV
resonance is dominated by s-wave capture, and the gamma-ray emission
should be not far from isotropic. However, a minor contribution from d-wave
capture cannot be excluded. In a simple potential model, the ratio of
Coulomb barrier transmission factors between d-wave and s-wave proton
capture  on $^{13}$C is about 2.5$\cdot$10$^{-3}$ at E$_{p}$ = 550 keV, 
which could lead to slight anisotropies of the
gamma-ray emission, e.g. up to 5\% for the 8062 keV to ground-state
transition.  The angular distributions were measured at four angles for this 
resonance to verify the level of isotropy. 

The gamma-ray angular distributions of the 1747 keV resonance are a-priori
not isotropic due the p-wave and f-wave character of the proton capture.
Only data for a few of the gamma-ray angular distributions are 
availabale in the literature for this resonance \cite{Pro} \cite{Sie}. 
We decided therefore to accurately determine the angular
distributions of the strongest gamma-ray lines. Data with the detector
D02 were taken at six angles between 0$^{\circ}$ and 90$^{\circ}$ while 
the 45$^{\circ}$ data from the two monitor detectors were also used for 
the determination of the angular distribution.

Normalization of the different efficiency curves of P2F, D03 and D02
(0$^{\circ}$) with respect to each other
were done with the 898 keV and 1836 keV lines of the
$^{88}$Y source. For D02 at
15$^{\circ}$, 30$^{\circ}$ and 45$^{\circ}$ the 0$^{\circ}$ efficiency curve
was corrected for transmission to account for the additional
absorption of the gamma rays in the target-support structure due to the
non-normal incidence of the gamma rays. The
60$^{\circ}$ and 90$^{\circ}$ efficiency curves of D02 were normalized to
the 0$^{\circ}$ efficiency with the help of the monitor detectors, using
the 1779 keV line of the  $^{27}$Al(p,$\gamma$) reaction, whose angular
distribution is well defined. The product of accumulated beam
charge Q and target thickness $\rho$ during the different runs was monitored by
the D03 and P2F detectors, where the intensity of the strongest gamma-ray
lines was used for the normalization.

The uncertainty of the angular distribution data points is due to the
uncertainties from peak integration, including the Q$\rho$ normalization 
with the
monitor detectors, and an estimated uncertainty in detector 
efficiency with respect to the 0$^{\circ}$ efficiency. This last one
was taken to be 2.5\% at 15$^{\circ}$ and 30$^{\circ}$, 5\% at
45$^{\circ}$ and 60$^{\circ}$ and 10 \% at 90$^{\circ}$. 
These uncertainties were added quadratically.

Angular distributions for gamma-ray lines of the 550 keV resonance are shown
in fig~3. All angular distributions are essentially isotropic,
with anisotropies not exceeding five per cent, with a probable exception of
the 4113 keV line, where it may reach ten per cent. The angular
distributions $W(\theta)$ have been adjusted by Legendre polynomials
$P_l(cos\theta$) of even order:

\begin{equation}
W(\theta) ~ = ~ \sum_{l=0}^{l_{max}} ~ a_l ~P_l(cos\Theta)
\label{eq2}
\end{equation}

, where $l_{max}$ is the smaller of twice the spin of the emitting state
$J_i$ or twice the transition multipolarity L. For this resonance, all
observed transitions have $J_i$ $\le$ 1, such that $l_{max}$ $\le$ 2. We did
not include the usual attenuation coefficients $Q_l$ because they differ
less than three per thousand from unity in our geometry.
The angle-integrated line intensities P and the $a_2/a_0$ coefficient ratios
are presented in table~3.

Results for the 1747 keV resonance are presented in table~4 and selected
angular distributions are shown in fig.~4.
For this resonance  only three of the studied gamma-ray transitions
have a unique assignment for the transition multipolarity: E1 and E2 for the
1635 and 6858 keV lines, respectively, and the 2313 keV line emission is
isotropic because of $J_i$ = 0. The two 3$^+$ $\rightarrow$ 1$^+$
transitions were assumed to be predominantly E2 and their angular
distributions fitted with $l_{max}$ = 4. 

Among the other nine transitions, four could be E1-M2 mixtures 
($\Delta J$ = 1; change of parity) and five M1-E2 mixtures 
($\Delta J$ = 1; no change of parity). For these nine
transitions, we performed fits with both $l_{max}$ = 2 and $l_{max}$ = 4.
We found that only the angular distribution of the 7027 keV line required an
$a_4$ coefficient different from zero, the fit improving considerably with
$l_{max}$ = 4. The fit for the 3338 keV line yielded an $a_4$ coefficient
slightly different from zero with, however, no significant improvement of
the fit. For the other seven distributions the $a_4$ coefficient was
compatible with zero, the reduced $\chi^2$ improving generally for $l_{max}$
= 2. Assuming that M2 (resp. E2) transitions are negligible with respect to
E1 (resp. M1) for these transitions, we adopted $l_{max}$ = 2.

\section{Results and discussion}

\subsection{The E$_p$ = 550 keV resonance}

A comparison of our integrated line intensities with literature values,
calculated from published gamma-ray branching ratios of the involved
levels in $^{14}$N, is shown in fig.~5. The values of the
compilation of Firestone et al. \cite{Fir} (identical to that of 
Ajzenberg-Selove \cite{Ajz}) (fig. 5a), 
are clearly in disagreement with our data with
the exception of three strong lines, while a substantial improvement can be 
seen, when replacing the compilation branching ratios of the 8062 keV level 
with those of the 
recent  experiments of King et al. \cite{King} (fig. 5b) or 
Zeps et al. \cite{Zeps} (fig. 5c).
Then, all relative line intensities are compatible with ours, 
except for the 
very weak line at 3947 keV which is about twenty per cent stronger in the
other data sets. 

It seems that the branching ratios of the 8062 keV level in the
compilations are taken essentially from the experimental determination of
Renan et al. \cite{Ren}. It is not clear why their line intensities
differ significantly from ours and that of \cite{King} \cite{Zeps} for
several lines. This experiment used also Ge detectors  and 
the $^{27}$Al(p,$\gamma$) reaction for efficiency calibration. 
A possible explanation could be the fact that  the efficiency calibration 
was based not only on full-energy peaks and that
older data and other resonances of the $^{27}$Al(p,$\gamma$) reaction 
were used. 

We extracted therefore from our data new branching ratios of the 8062 keV
level. The intensities for the transitions to the 3948 keV and
the ground state have been taken directly from the corresponding relative
line intensities. The transition intensities to the 4915 and 5691 keV
levels have been taken from the 4914, 5690 and
3378 keV gamma-ray transitions which depopulate these levels.
The gamma-ray lines
corresponding to transitions from the 8062 keV level to the 2313 and
5106 keV levels were not
strong enough to obtain angular distributions. They were only extracted
from summed spectra of the two monitor detectors at 45$^{\circ}$. 
We added for their relative line intensities a systematic uncertainty of 
ten per cent to the  standard uncertainties 
due to possible anisotropies in the angular distributions. Our 8062 keV level 
branching ratios are presented in table~5 together with previous
data. Branching ratios for the levels at 3948 keV and 5691 keV could also be
extracted. They are in agreement with the compilation values
but are less precise than these and are therefore not presented.

\subsection{The E$_p$ = 1747 keV resonance}

Some angular distributions for this resonance have been published before.
The most accurate and extensive data can be found in Prosser et al.
\cite{Pro} and Sievers et al. \cite{Sie}, who
measured the angular distributions of the 9169, 7027, 6857, 6446, 2726,
and 2143 keV lines. Prosser et al. published the results of Legendre 
polynomial fits to the angular distributions. They are
in good agreement with ours, except for the 7027 keV line, where 
their  a$_2$/a$_0$ and a$_4$/a$_0$ coefficients differ from ours 
sligthly. This may be explained by the fact that
Prosser et al. used NaI detectors, where the 7027 and 6857 keV lines were
superposed and sitting on the tail of the 9169 keV line complex. Good to 
excellent agreement with the measured angular distributions of 
Sievers et al. is observed.  

The comparison of our line intensities with the values calculated from the
published gamma-ray branching ratios  \cite{Fir} in fig.~6
shows good agreement, with the notable exception of the 728, 3338 and
5106 keV lines. Most striking is the
difference for the 728 keV line where our intensity is
about a factor of two lower than the literature value. This line corresponds
to the only transition depopulating the 5834 keV level (see fig. 2). 
A small part of the
difference can be explained by the fact that we find slightly lower 
intensity of the 3338 keV line in our
experiment, which populates the 5834 keV level. 

The only other transition indicated in the compilation
which populates this level, 6446 keV $\rightarrow$ 5834 keV, 
produces a 612 keV line
which should have an intensity of about 0.4\% with respect to the 9169 keV
line.
However, the 612 keV line is Doppler broadened and very weak in our spectra
and in addition close to
the 609 keV background line from the $^{226}$Ra decay chain. We could only
extract an upper limit of 0.1\% with respect to the 9169 keV line. The sum
of both transitions populating the 5834 keV level in our experiment is 
compatible with our extracted intensity of the 728 keV transition
depopulating this level.
We conclude, that the most probable explanation of the 
disagreement is the 6446 keV $\rightarrow$ 5834 keV transition, whose branching
ratio is at least a factor of four smaller than given in the compilations.   

New branching ratios for the 9172 keV and the 6446 keV level are presented in
table~5 together with values of the compilations. For both levels,
we have extracted angular distributions of all depopulating transitions 
listed in the compilations with the exception of the very weak
6446 keV $\rightarrow$ 5834 keV transition, and our branching ratios were 
taken directly from the 
respective angle-integrated gamma-ray line intensities.

\subsection{Conclusion}

The E$_p$ = 550 keV and 1747 keV resonances of the
 $^{13}$C(p,$\gamma$)$^{14}$N reaction provide an interesting alternative to
the widely used  E$_p$ = 992 keV resonance of the 
$^{27}$Al(p,$\gamma$)$^{28}$Si reaction for energy and
efficiency calibration of gamma-ray detectors in the energy range from about
1.6 MeV to 9 MeV. Relative line intensities of eight gamma-ray lines at five
angles with respect to the proton beam for the
550 keV resonance and of fourteen lines at six different angles
for the 1747 keV resonance have been obtained in this study. Uncertainties
in relative line intensities
are typically of the order of five per cent for a dozen strong lines 
and even slighter better at
45$^{\circ}$, allowing to use these data directly for precise efficiency
calibrations at the chosen specific angles. Additionally, gamma-ray angular
distributions have been obtained for all twenty-two gamma-ray lines to
permit efficiency calibrations at any angle and 
allowing the extraction of gamma-ray branching ratios for several levels of
$^{14}$N. We present new branching ratios for the 6446, 8062 and 
9172 keV levels. In particular, we propose new branching ratios 
for the 8062 keV level which are consistent with two other recent 
experiments, but differ considerably from the values of the compilations
\cite{Fir} \cite{Ajz}.

\newpage

\begin{table} [t]
\begin{tabular}  {cccccc} \hline
E$_{\gamma}$ (keV) & 0$^{\circ}$ & 30$^{\circ}$ & 45$^{\circ}$ &
60$^{\circ}$ & 90$^{\circ}$ \\
\hline
8060 &  100  & 100 &  100 & 100 &  100 \\
5690 &  1.94$\pm$0.14 & 2.12$\pm$0.16 & 2.05$\pm$0.13 & 2.19$\pm$0.18 &
2.00$\pm$0.18 \\
4914 &  2.53$\pm$0.17 & 2.66$\pm$0.21 & 2.60$\pm$0.13 & 2.57$\pm$0.20 &
2.56$\pm$0.23 \\
4113 &  16.62$\pm$0.87 & 16.51$\pm$0.89 & 16.32$\pm$0.66 & 16.25$\pm$0.87 &
16.24$\pm$0.90 \\
3947 &  0.52$\pm$0.07 & 0.50$\pm$0.09 & 0.52$\pm$0.06 & 0.55$\pm$0.11 &
0.53$\pm$0.12 \\
3378 &  3.82$\pm$0.21 & 3.72$\pm$0.21 & 3.79$\pm$0.18 & 3.79$\pm$0.22 &
4.03$\pm$0.24 \\
2313 &  21.93$\pm$1.11 & 21.67$\pm$1.11 & 22.20$\pm$0.73 & 22.52$\pm$1.16 &
23.78$\pm$1.24 \\
1635 &  15.53$\pm$0.79 & 15.33$\pm$0.79 & 15.94$\pm$0.50 & 16.67$\pm$0.86 &
17.21$\pm$0.90 \\  
\hline
\end{tabular}
\label{table1}
\caption{\scriptsize
 Relative line intensities of the E$_p$ = 550 keV
resonance with respect to the 8060 keV line at five different angles. } 
\end{table}

\begin{table} [t]
\begin{tabular}  {ccccccc} \hline
E$_{\gamma}$ (keV) & 0$^{\circ}$  & 15$^{\circ}$ & 30$^{\circ}$ &
45$^{\circ}$  & 60$^{\circ}$ & 90$^{\circ}$  \\
\hline
9169 & 100 & 100 & 100 & 100 & 100 & 100 \\
7027 & 1.61$\pm$0.14& 2.38$\pm$0.19& 3.18$\pm$0.20& 3.75$\pm$0.13& 3.45$\pm$0.20&
3.27$\pm$0.20 \\
6857 & 1.06$\pm$0.24& 1.46$\pm$0.14& 2.01$\pm$0.14& 2.02$\pm$0.08& 1.20$\pm$0.10&
0.40$\pm$0.19 \\
6445 & 18.05$\pm$0.94  & 17.56$\pm$1.02  & 14.41$\pm$0.75& 10.22$\pm$0.32&
6.93$\pm$0.37& 4.00$\pm$0.30 \\
5105 & 1.73$\pm$0.27& 1.94$\pm$0.57&  1.38$\pm$0.23& 1.01$\pm$0.06& 0.88$\pm$0.20
& 0.86$\pm$0.19 \\
3480 & 0.36$\pm$0.11& 0.35$\pm$0.08& 0.35$\pm$0.06& 0.434$\pm$0.024& 0.49$\pm$0.13&
0.41$\pm$0.07 \\
3338 & 0.75$\pm$0.12& 0.81$\pm$0.10& 0.74$\pm$0.05& 0.571$\pm$0.025& 0.46$\pm$0.05
& 0.41$\pm$0.06 \\
2726 & 17.64$\pm$0.89  & 16.92$\pm$0.86& 13.62$\pm$0.69& 11.50$\pm$0.33&
9.80$\pm$0.50& 8.83$\pm$0.45 \\
2498 & 5.28$\pm$0.27& 5.08$\pm$0.27& 3.81$\pm$0.20& 2.89$\pm$0.09& 1.95$\pm$0.10&
1.07$\pm$0.06 \\
2313 & 7.46$\pm$0.38& 7.16$\pm$0.37& 5.67$\pm$0.29& 4.52$\pm$0.13& 3.71$\pm$0.19&
3.39$\pm$0.18 \\
2143 & 9.80$\pm$0.50& 9.21$\pm$0.47& 6.30$\pm$0.32& 4.55$\pm$0.13& 3.17$\pm$0.16&
2.26$\pm$0.12 \\
1635 & 2.77$\pm$0.20& 2.72$\pm$0.18& 2.40$\pm$0.16& 2.33$\pm$0.08&
2.19$\pm$0.13 & 2.11$\pm$0.13 \\
1340 & 0.82$\pm$0.07& 0.90$\pm$0.06& 0.73$\pm$0.06& 0.697$\pm$0.025& 0.64$\pm$0.05&
0.64$\pm$0.06 \\
728  & 0.48$\pm$0.06& 0.52$\pm$0.07  & 0.39$\pm$0.05& 0.428$\pm$0.027& 
0.46$\pm$0.06 & 0.36$\pm$0.06 \\
\hline
\end{tabular}
\label{table2}
\caption{\scriptsize Relative line intensities of the E$_p$ = 1747 keV
resonance with respect to the 9169 keV line at six different angles. } 
\end{table}

\begin{table} [t]
\begin{tabular}  {cccccc} \hline
E$_{\gamma}$ (keV) & E$_i$ $\rightarrow$ E$_f$ & J$_i$ $\rightarrow$ J$_f$ & 
P & a$_2$/a$_0$ &  $\chi^2_{red}$ \\ 
\hline
8060 &  8062 $\rightarrow$ 0 & 1$^-$ $\rightarrow$ 1$^+$ & 100$\pm$3.3 &
0.037$\pm$0.042 & 0.19 \\
5690 &  5691 $\rightarrow$ 0 & 1$^-$ $\rightarrow$ 1$^+$ & 2.08$\pm$0.15
&  -0.0096$\pm$0.068 &  1.05 \\
4914 &  4915 $\rightarrow$ 0 & 0$^-$ $\rightarrow$ 1$^+$ & 2.63$\pm$0.16
& & 0.29 \\
4113 &  8062 $\rightarrow$ 3948 & 1$^-$ $\rightarrow$ 1$^+$ & 16.3$\pm$1.0
& 0.056$\pm$0.045 &  0.14 \\
3947 &  3948 $\rightarrow$ 0 & 1$^+$ $\rightarrow$ 1$^+$ & 0.522$\pm$0.064
&  0.017$\pm$0.16 &  0.06 \\
3378 &  5691 $\rightarrow$ 2313 & 1$^-$ $\rightarrow$ 0$^+$ & 3.79$\pm$0.24
&  0.034$\pm$0.048 &  0.08 \\
2313 &  2313 $\rightarrow$ 0 & 0$^+$ $\rightarrow$ 1$^+$ & 22.5$\pm$1.2
& &  0.06 \\
1635 &  3948 $\rightarrow$ 2313 & 1$^+$ $\rightarrow$ 0$^+$ & 16.2$\pm$1.0 
& -0.019$\pm$0.040 &  0.33 \\
\hline
\end{tabular}
\label{table3}
\caption{\scriptsize  Angle-integrated relative line intensity P
and Legendre polynomial coefficients from fits of the gamma-ray angular
distributions  of the E$_p$ = 550 keV resonance of the $^{13}$C(p,$\gamma$)
reaction. The uncertainties given to P include a contribution due to
$a_0$ and an additional contribution of five per cent which accounts
for the uncertainty of the detector efficiency curve for lines
other than the 8060 keV line.} 
\end{table}

\begin{table} [t]
\begin{tabular}  {ccccccc} \hline
E$_{\gamma}$ (keV) & E$_i$ $\rightarrow$ E$_f$ & J$_i$  $\rightarrow$ J$_f$ & 
P &  a$_2$/a$_0$ & a$_4$/a$_0$ & $\chi^2_{red}$ \\
\hline
9169 & 9172 $\rightarrow$ 0 & 2$^+$ $\rightarrow$ 1$^+$ & 100$\pm$2.5 &
-0.466$\pm$0.029 & & 0.60 \\
7027 & 7029 $\rightarrow$ 0 & 2$^+$ $\rightarrow$ 1$^+$ & 3.35$\pm$0.22 &
-0.491$\pm$0.080 & -0.248$\pm$0.050 & 1.31 \\
6858 & 9172 $\rightarrow$ 2313 & 2$^+$ $\rightarrow$ 0$^+$ & 1.12$\pm$0.10
& 0.509$\pm$0.138 & -1.035$\pm$0.134 & 1.41 \\
6445 & 6446 $\rightarrow$ 0 & 3$^+$ $\rightarrow$ 1$^+$ & 7.46$\pm$0.37 &
0.560$\pm$0.060 & -0.259$\pm$0.045 & 0.37 \\
5105 & 5106 $\rightarrow$ 0 & 2$^-$ $\rightarrow$ 1$^+$ & 0.92$\pm$0.10 &
0.012$\pm$0.162 &  & 0.23 \\
3480 & 9172 $\rightarrow$ 5691 & 2$^+$ $\rightarrow$ 1$^-$ & 0.429$\pm$0.045 &
-0.58$\pm$0.14 & & 0.67 \\
3338 & 9172 $\rightarrow$ 5834 & 2$^+$ $\rightarrow$ 3$^-$ & 0.514$\pm$0.039 &
-0.065$\pm$0.090 & & 0.75 \\
(3338) &  &   & 0.496$\pm$0.040 & 0.034$\pm$0.116 & -0.165$\pm$0.109 & 0.22 \\
2726 & 9172 $\rightarrow$ 6446 & 2$^+$ $\rightarrow$ 3$^+$ & 10.34$\pm$0.59 &
-0.084$\pm$0.031 & & 1.36 \\
2498 & 6446 $\rightarrow$ 3948 & 3$^+$ $\rightarrow$ 1$^+$ & 2.07$\pm$0.12 &
0.567$\pm$0.054 & -0.198$\pm$0.043 & 2.83 \\
2313 & 2313 $\rightarrow$ 0 & 0$^+$ $\rightarrow$ 1$^+$ & 4.01$\pm$0.21 &
 & & 1.12 \\
2143 & 9172 $\rightarrow$ 7029 & 2$^+$ $\rightarrow$ 2$^+$ & 3.56$\pm$0.20 &
0.474$\pm$0.036 & & 2.70 \\
1635 & 3948 $\rightarrow$ 2313 & 1$^+$ $\rightarrow$ 0$^+$ & 2.24$\pm$0.12 &
-0.349$\pm$0.025 & & 0.67 \\
1340 & 6446 $\rightarrow$ 5106 & 3$^+$ $\rightarrow$ 2$^-$ & 0.654$\pm$0.044 &
-0.281$\pm$0.058 & & 1.05 \\
728 & 5834 $\rightarrow$ 5106 & 3$^-$ $\rightarrow$ 2$^-$ & 0.412$\pm$0.047 &
-0.387$\pm$0.079 & & 1.72 \\

\hline
\end{tabular}
\label{table4}
\caption{\scriptsize  Angle-integrated relative line intensity P
and Legendre polynomial coefficients from fits of the gamma-ray angular
distributions  of the E$_p$ = 1747 keV resonance of the $^{13}$C(p,$\gamma$)
reaction. The uncertainty given to P include a contribution due to $a_0$
and an additional contribution of five per cent which accounts for
the uncertainty of the detector efficiency curve for lines
other than the 9169 keV line.} 
\end{table}

\begin{table} [t]
\begin{tabular}  {cccccc} \hline
level  & transition  & & branching ratios & & \\
E$_i$ (keV) & $\rightarrow$ E$_f$ (keV) &  this work & King et al.
\cite{King} & Zeps et
al. \cite{Zeps} & compilations \cite{Fir} \cite{Ajz} \\
\hline
8062 & & & & &  \\
 & 0 & 100$\pm$0.7  & 100$\pm$5.8 & 100$\pm$0.5 & 100$\pm$0.8 \\
 & 2313 & 2.19$\pm$0.33 & 2.39$\pm$0.35 & 1.5$\pm$0.9 & 1.74$\pm$0.18 \\
 & 3948 & 16.3$\pm$1.0  & 16.9$\pm$1.4 & 15.9$\pm$0.8 & 15.8$\pm$0.5 \\
 & 4915 & 2.63$\pm$0.20 & 2.89$\pm$0.30 & 2.1$\pm$0.5 & 2.32$\pm$0.18 \\
 & 5106 & 0.73$\pm$0.14 & 0.83$\pm$0.16 & 1.0$\pm$0.5 & 0.31$\pm$0.18 \\
 & 5691 & 5.87$\pm$0.28 & 6.7$\pm$1.4 & 5.7$\pm$0.5 & 4.4$\pm$0.5 \\
9172 & & & & & \\
 & 0 & 100$\pm$0.9 & & & 100$\pm$1.2 \\
 & 2313 & 1.12$\pm$0.10 & & & 1.00$\pm$0.10 \\
 & 5691 & 0.43$\pm$0.05 & & & 0.58$\pm$0.12 \\
 & 5834 & 0.51$\pm$0.04 & & & 0.72$\pm$0.10 \\
 & 6446 & 10.34$\pm$0.64 & & & 10.4$\pm$1.0 \\
 & 7029 & 3.56$\pm$0.22 & & & 3.7$\pm$0.4 \\
6446 & & & & & \\
 & 0 & 100$\pm$2.7 & & & 100$\pm$2.1 \\
 & 3948 & 27.7$\pm$2.1 & & & 28.1$\pm$1.4 \\
 & 5106 & 8.8$\pm$0.7 & & & 9.3$\pm$0.9 \\
 & 5834 & $<$1.3 & & & 5.3$\pm$0.9 \\
\hline
\end{tabular}
\label{table5}
\caption{ \scriptsize
Branching ratios of the 8062 keV, 9172 keV and 6446 keV levels of
$^{14}$N. }
\end{table}

\newpage

\begin{figure}
\includegraphics*[0cm,1cm][12cm,10cm]{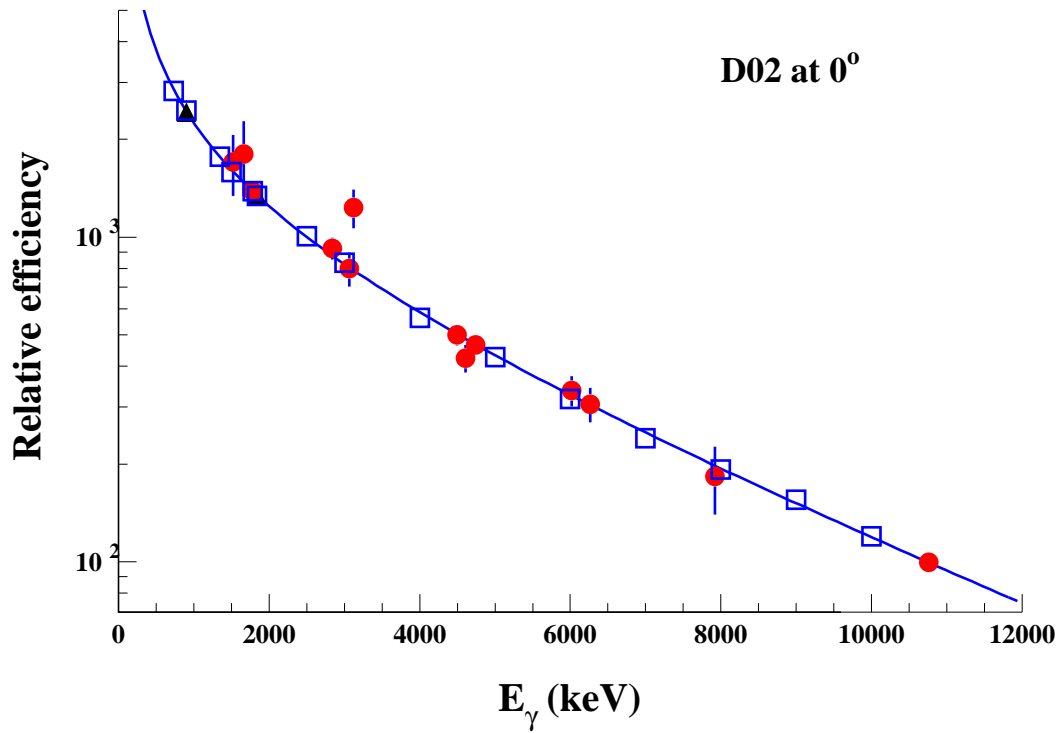}
\caption{\scriptsize
Relative photo peak efficiency of the detector D02 at 0$^{\circ}$,
determined with the $^{27}$Al(p,$\gamma$)$^{28}$Si reaction
and the $^{88}$Y source.  
Experimental data are the filled symbols (black triangles
for the  $^{88}$Y source and red circles for the (p,$\gamma$) reaction), the
power-law fit to the data is shown as a continuous blue line. The open
squares
represent the relative efficiencies resulting from GEANT simulations of
this detector. The absolute value of the simulated efficiencies have been
normalized to the experimental data at the 1779 keV line. No reason for
the strong deviation of the 3123.7 keV data point from the other data 
could be found. Its influence on the fit results is, however, not very 
strong and it was decided to keep it in the fits.}
\end{figure}

\newpage

\begin{figure}
\includegraphics*[0cm,0cm][12cm,11cm]{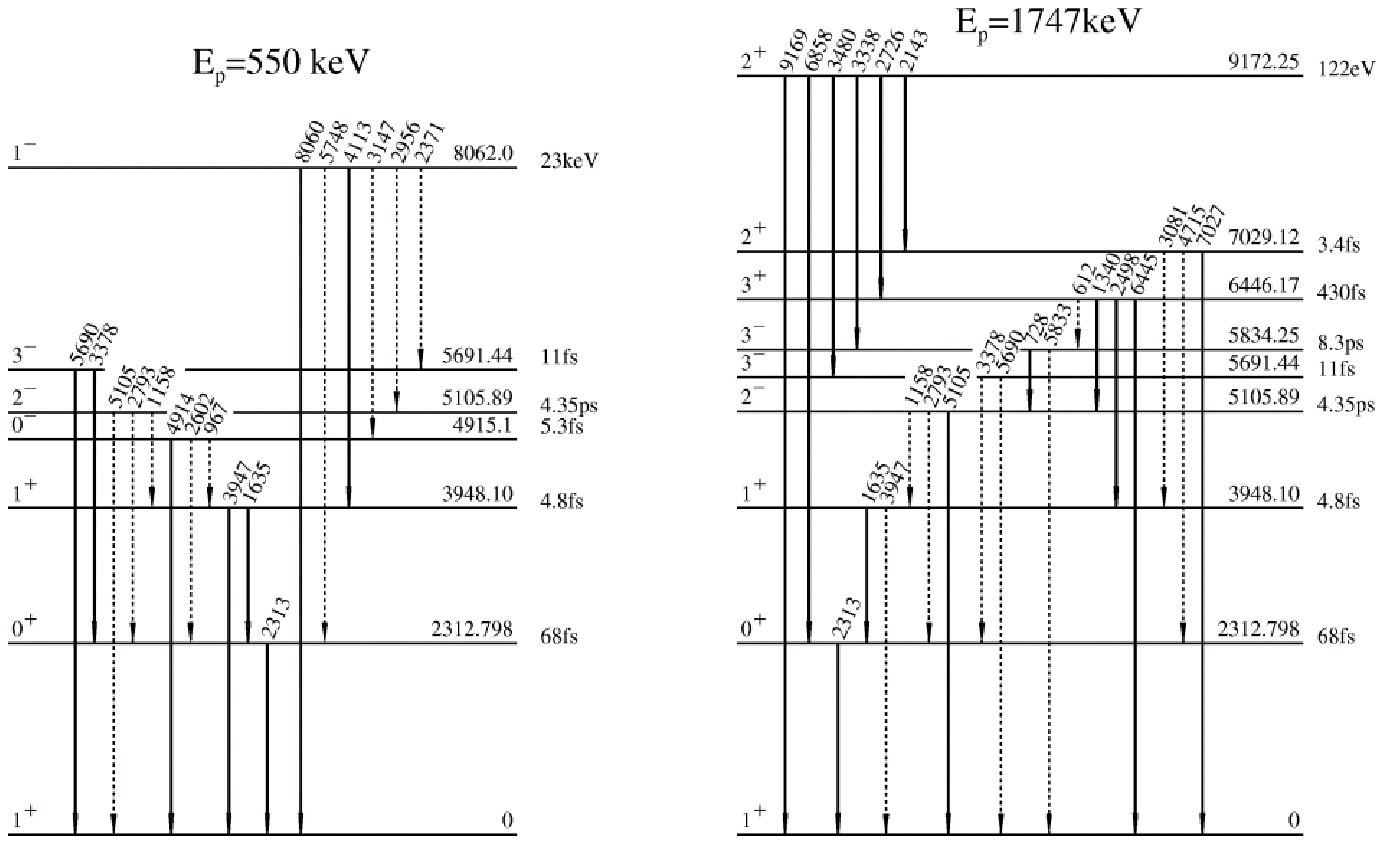}
\caption{\scriptsize
Decay schemes of the 9172 keV, 2$^+$ level of $^{14}$N, 
which
corresponds to the E$_p$ = 1747 keV resonance (right) and of the
8062 keV, 1$^-$ level of $^{14}$N, 
which corresponds to the E$_p$ = 550 keV resonance  (left) of the 
$^{13}$C(p,$\gamma$) reaction. Only levels which are involved in the decay
of the respective resonance are plotted. Transitions which have been studied
in this work are indicated by full arrows; other known 
transitions are indicated
by broken arrows. Spins, energies and half lifes are taken from
the compilation of Firestone et al. \cite{Fir}. }
\end{figure}

\newpage

\begin{figure}
\includegraphics*[0cm,1cm][12cm,14cm]{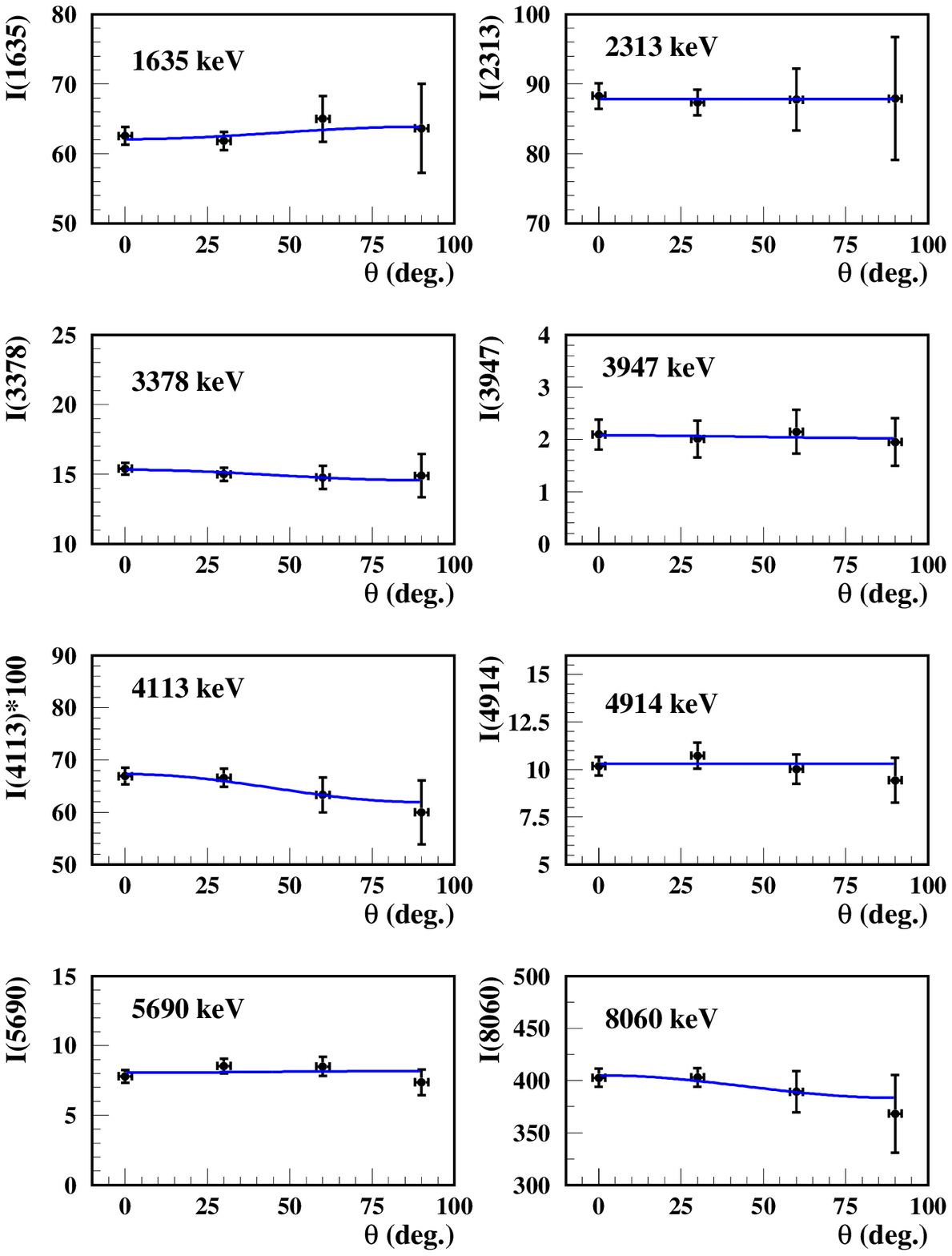}
\caption{\scriptsize
Measured gamma-ray angular distributions of the E$_p$ = 550 keV 
resonance together with the Legendre polynomial fits to the data.}
\end{figure}

\newpage

\begin{figure}
\includegraphics*[0cm,1cm][12cm,14cm]{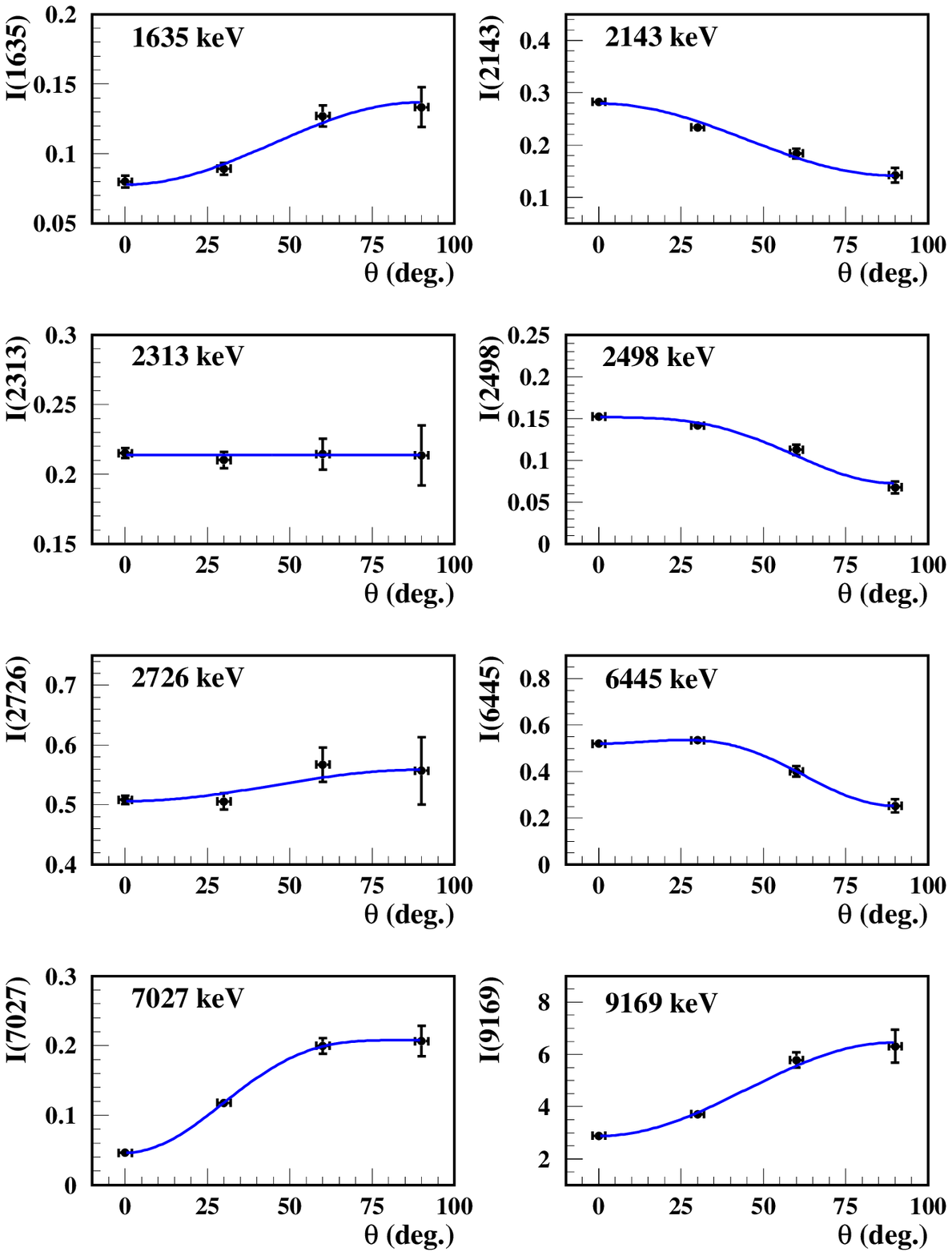}
\caption{\scriptsize
Measured gamma-ray angular distributions of the E$_p$ = 1747 keV
resonance together with the Legendre polynomial fits to the data.}
\end{figure}

\newpage

\begin{figure}
\includegraphics*[0cm,1cm][12cm,14cm]{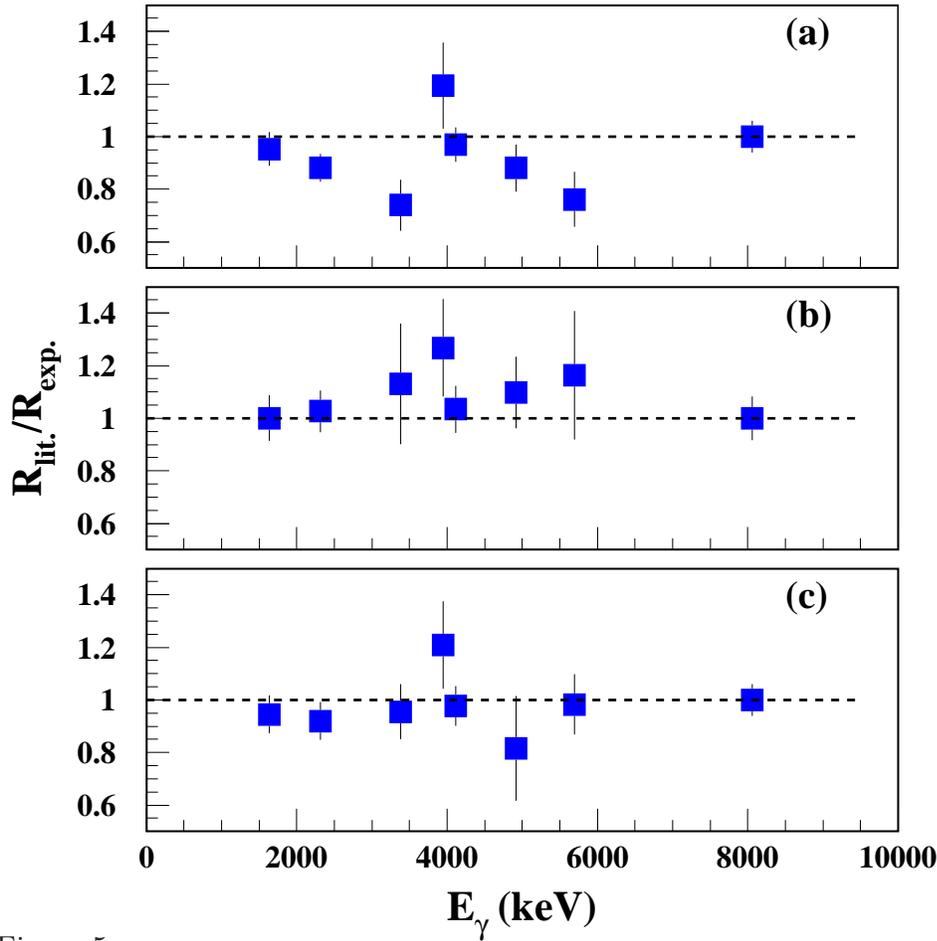}
\caption{\scriptsize
Ratios of the extracted relative gamma-ray line intensities
R$_{exp.}$ for the E$_p$ = 550 keV resonance
from this experiment with literature values R$_{lit.}$ with
R$_{lit.}$(8060 keV)~/~R$_{exp.}$(8060 keV)~$\equiv$1. The different
line intensities from literature have been calculated from the gamma-ray
branching ratios of $^{14}$N levels, taken from: 
(a) the compilations of Firestone et al. 
and Ajzenberg-Selove \cite{Fir} \cite{Ajz}
(b) the same compilations, except branching ratios of the 8062 keV
level from the experiment of King et al. \cite{King} {c}  the same
compilations, except branching ratios of the 8062 keV
level from the experiment of Zeps et al. \cite{Zeps}. }
\end{figure}
\vspace{2 cm}

\begin{figure}
\includegraphics*[0cm,1cm][12cm,14cm]{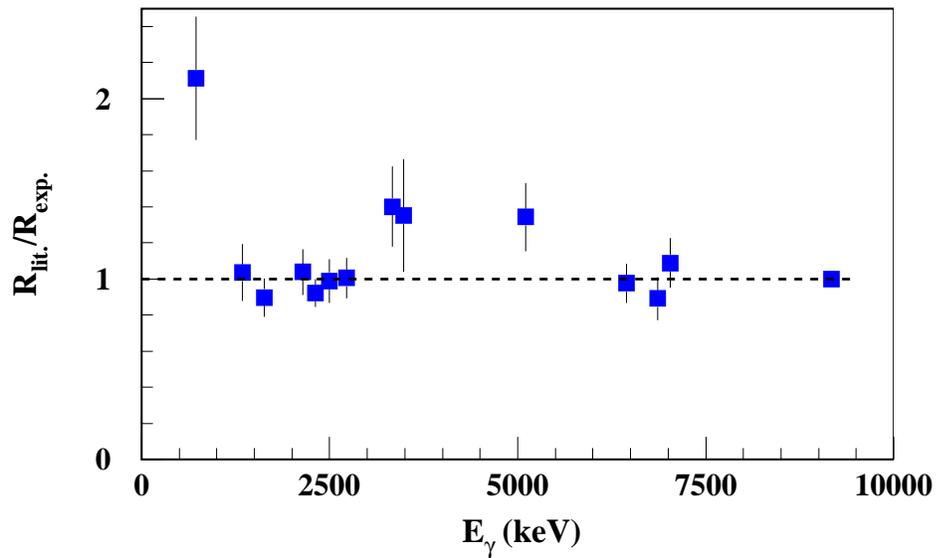}
\caption{\scriptsize
Ratio of the extracted relative gamma-ray line intensities
R$_{exp.}$ for the E$_p$ = 1747 keV resonance
from this experiment with literature values R$_{lit.}$ with
R$_{lit.}$(9169 keV)~/~R$_{exp.}$(9169 keV)~$\equiv$1. Literature values have
been calculated from the branching ratios of $^{14}$N levels given in
the compilations  \cite{Fir} \cite{Ajz}.}
\end{figure}

\end{document}